\documentclass{article}

\usepackage[final,nonatbib]{neurips_2024}

\usepackage[utf8]{inputenc} % allow utf-8 input
\usepackage[T1]{fontenc}    % use 8-bit T1 fonts
\usepackage{hyperref}       % hyperlinks
\usepackage{url}            % simple URL typesetting
\usepackage{booktabs}       % professional-quality tables
\usepackage{amsfonts}       % blackboard math symbols
\usepackage{nicefrac}       % compact symbols for 1/2, etc.
\usepackage{microtype}      % microtypography
\usepackage{xcolor}         % colors
\usepackage{graphicx}
\usepackage{float}
\usepackage{multirow}
\usepackage{amsmath}
\usepackage{subcaption}

\title{Transfer Learning in Materials Informatics: structure-property relationships through minimal but highly informative multimodal input }

\author{
    Dario Massa\\
    IDEAS-NCBR Chmielna 69 00-801 Warsaw, Poland\\
    NOMATEN Center of Excellence in Multifunctional Materials, \\
    ul. Andrzeja Sołtana, 05-400 Otwock, Świerk, Poland\\
    dario.massa@ncbj.gov.pl\\
    \And
    Grzegorz Kaszuba\\
    IDEAS-NCBR Chmielna 69 00-801 Warsaw, 
    Poland\\
    Poznań University of Technology, Poland\\
    \And
    Stefanos Papanikolaou\\
    NOMATEN Center of Excellence in Multifunctional Materials,\\
    ul. Andrzeja Sołtana, 05-400 Otwock, Świerk, Poland\\
    \And
    Piotr Sankowski\\
    IDEAS-NCBR Chmielna 69 00-801 Warsaw, Poland\\
    MIM Solutions, Poland\\
    piotr.sankowski@mim.ai\\}
\begin{document}

\maketitle

\begin{abstract}
In this work we propose simple, effective and computationally efficient transfer learning approaches for structure-property relation predictions in the context of materials, with highly informative input from different modalities. As materials properties stand from their electronic structure, representations are extracted directly from datasets of electronic charge density profile images using Neural Networks. We demonstrate transferability of the existing pre-trained Convolutional Neural Networks and Large Language Models knowledge to physics domain data, exploring a wide set of compositions for the regression of energetics- or structure- related properties, and the role of semantic crystallographic information in the context of multimodal approaches. We test the applicability of the CLIP multimodal model, and employ as well a training protocol for building a more interpretable and versatile stacked custom solution from different pre-trained modalities.  The study offers a promising avenue for enhancing the effectiveness of descriptor identification in physical systems, shedding light on the power of multimodal transfer learning for materials property prediction. Instead of using the well-established GNN-based approaches, we explore the transfer learning of image- and text-based architectures, which can impact decision making for new low-cost AI methods in the field of Materials and Chemoinformatics.
\end{abstract}

\section{Introduction}

%%MI
Machine Learning (ML)~\cite{ML1,ML2} has initiated a remarkable transformation in Materials Science, with the emergence of Materials Informatics (MI) playing a pivotal role in pushing the boundaries of innovation in the context of multiple technological research challenges ~\cite{MI_2,MI_4,MI_5,CcaptureML_2,CcaptureML_1,organic_2,organic,electrocatML_2,electrocatML_1}. A key aspect is the identification of efficient descriptors, capable of capturing the essential features of complex physical systems. In the realm of materials simulations, a wide variety of descriptors have been proposed, both based on translationally and rotationally invariant functions of atomic coordinates ~\cite{CoulombMat,ACSFs,SPRINT,SOAP,globalmin}, or involving fixed-length feature vectors of atomic or electronic properties~\cite{fixlen_feat_1,fixlen_feat_2,fixlen_feat_3}, as well as more advanced feature extraction approaches through Graph Convolutional Neural Networks (GCNNs)~\cite{GNNcry1,GNNcry3,GNNsurf1,GNNmol_1,GNNmol_2,GNNmol_3,GNNmol_4}. Wide open-source databases~\cite{DB1,DB2,DB3,DB4}, originating from high-throughput~\cite{HighTrough_1,HighTrough_2} ab-initio Density-Functional-Theory (DFT)~\cite{dft1_general,dft2_general} calculations, represent the data source for the development of such models and MI in general. 

In real-life research settings, it is very common to come across low-quantity but highly informative data about the systems of interest: in the field of Molecular Science, an example can be found in molecular orbitals, encoding reactivity and optical properties; in the field of Materials Science, for example, phonon dispersion relations encode information on thermal conductivity. Recent works have demonstrated the power of deploying Electron Charge Density (ECD) profiles in ML for unsupervised studies of defects-related effects in metals~\cite{massa2023alloy}, or prediction of density fields in unseen samples via graph-based models~\cite{koker2024higher}. The interest in such quantity lies in its encoding of the necessary information regarding the properties of the system -- in terms of constitutive atoms and their interactions - regardless of dimensionality, stability or geometry. In this view, the ECD profiles represent an example of highly informative data, which could be efficiently exploited for structure-property relations also with the simplest available solutions. 

In this work, we aim to propose to utilize the computational efficiency of multimodal transfer learning solutions to predict materials properties from ECD profiles \cite{torrey2010transfer}. We build an image and text dataset based on the Materials Project (MP)~\cite{DB2}, which could stand as a representative example of a relatively small-sized but very diverse problem to be learned. We successfully test the transferability of the known multimodal Contrastive Language-Image Pre-training (CLIP) model from OpenAI~\cite{clip}. Furthemore, we employ a protocol for building a custom, more interpretable and versatile stacked multimodal solution starting from single pre-trained modalities. The proposed setting stands as an impactful solution involving minimal information and computational effort which could lay the basis for novel low-cost AI models for inverse materials design and databases in the field of Materials and Cheminformatics. 

\section{Methods}

\subsection{Multimodal dataset}

\begin{figure}[t]
    \centering
    \includegraphics[width=0.99\linewidth]{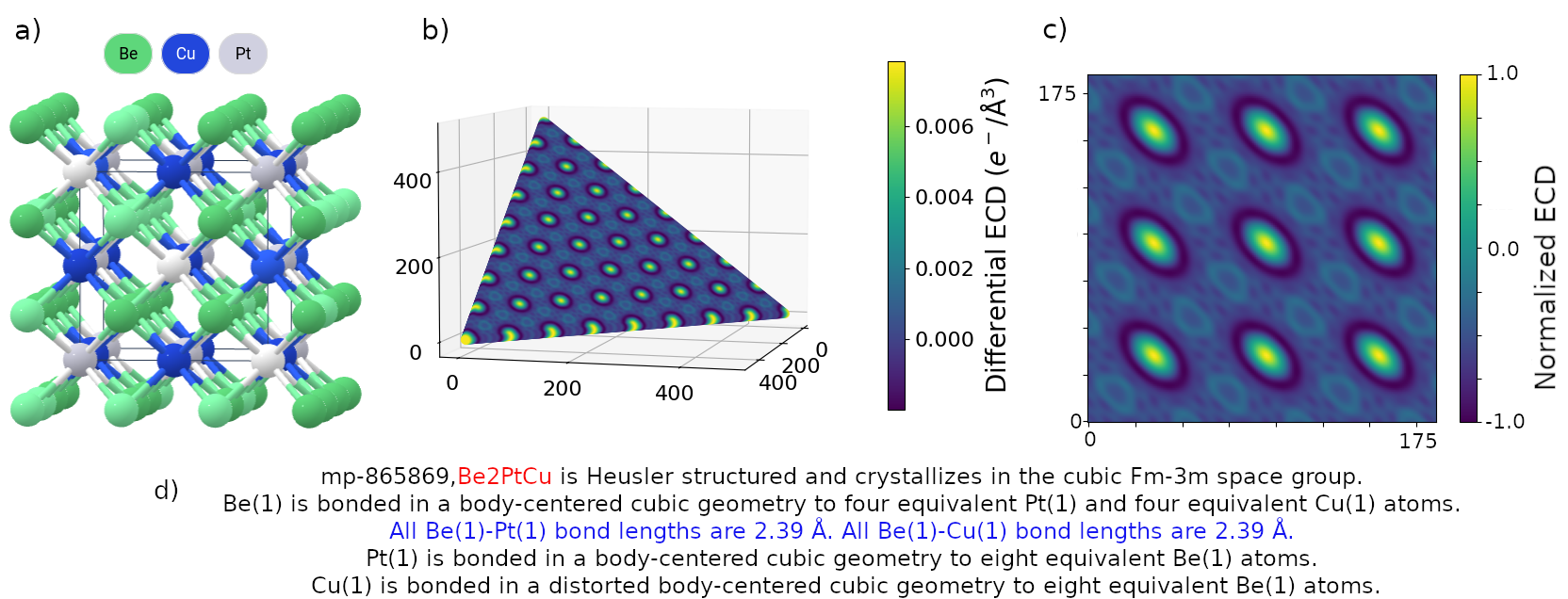}
    \caption{Visualization of the creation steps of the multimodal dataset. a) A bulk crystal is considered together with its electron charge density field, its target properties and its structure file (.cif). b) The differential electron charge density field is sliced along the maximum atomic density plane from an 8x8x8 repetition of the sample. c) A squared focus from the slice is extracted, and properly normalized. d) The crystallographic textual information about the sample is produced. During experiments, we also consider fractions of text as: 'Formula' (only the composition name), 'Keywords' (the composition name and the bond lengths information), and 'All' (the entire text).}
    \label{fig:dataset}
\end{figure}

We build our multimodal dataset as a subset of the Materials Project ~\cite{DB2}, restricting it to stable face-centered cubic (FCC) crystals. Their MP-IDs, the associated target properties and ECD fields are collected, resulting in $781$ samples. Data-cleaning is performed to filter out outliers in the distribution of maxima and minima of the collected ECD fields, through definition of acceptance boundaries with the interquartile range (IQR), resulting in $592$ samples. As shown in Fig.(\ref{fig:datadistrib}) of the Appendix, the majority of the collected samples is represented by ternary crystals, and their quantum-level characterization is very demanding from a computational point of view.
We use the dedicated Py-Rho~\cite{shen2021representationindependent} to unify the grid dimensions to $60\times60\times60$. To further simplify the problem without losing the descriptive power of the data, we slice the three-dimensional data along specific bi-dimensional plane, characterized by the highest atomic density. Having restricted ourselves to FCC crystals, such plane corresponds to the $\langle111\rangle$ Miller indices. As shown in Fig.(\ref{fig:dataset}), we perform the slicing in a supercell made by an $8\times8\times8$ repetition of the unit-cell field data, and focus the extraction of a squared image from such slice. Even though both total and differential ECDs are available, we restrict the dataset on the latter type for the enhanced variability within and among samples originating from the highlighting of interatomic interactions, rather than local atomic electron densities. A comparison between the slices along the z-axis of the total and differential ECD fields is presented in Appendix in Fig.(\ref{fig:zlices}). 
For regression experiments, we target energetics, with the formation ($\text{E}_{\text{F}}$) and Fermi energies ($\text{E}_{\text{fermi}}$), as well as a structural property, the bulk modulus ($\text{K}_{\text{vrh}}$). 
Before application of the models, the target properties undergo standardization, and the ECD isoline values a z-score normalization and rescaling in the $[-1;1]$ range: this allows to preserve peculiar patterns of zero-level oscillations across electron accumulation and depletion isolines.
Finally, with the help of the Robocrystallographer~\cite{Robocrys}, a tool capable of generating concise text descriptions of crystalline structures by analyzing their symmetry, local environment, and extended connectivity from their Pymatgen structures, we build a set of text attributes associated to each MP-ID, an example of which is reported in Fig.~(\ref{fig:dataset}).

\subsection{Multimodal models}

We use two foundational models to incorporate the textual and image information about the crystal structures, resulting in a stacked custom multimodal model. The schematic representation of the architecture is reported in Appendix in Fig.(\ref{fig:scheme}) To process the images, we use InceptionV3, a model pre-trained on ImageNet. For text, we use Roberta, trained on language modeling task. We use the same procedure to adapt either of the models. First, we remove model's classification head and replace it with a readout network composed of 4 dense layers, whose width gradually decreases from pre-trained model's latent dimension to 1 output channel. We first fit the new dense readout, to then unfreeze the rest of the model.
After training the image- and text-based property regressors, we load them both to create a multimodal model. We remove the final layer of both readout dense networks. We concatenate the outputs of penultimate layers of our models, each having 256 channels, and create two new readout layers. Once again, we first fit the newly created layers to then unfreeze the remainder of the model. All dense readout layers, apart from the ultimate single-channel layer, are followed by ReLU activation, BatchNorm and dropout. To contrast this approach, we also conduct experiments with CLIP, a transformer-based model pre-trained on a text-image multimodal task. We add a single linear layer on top of the pre-trained model to get the regression output.

\section{Experiments}
First, we verify the transferability of the CLIP multimodal model to our physics domain problem. To do so, we use $85\%$ of the dataset for cross-validation experiments with five folds, and keep the remaining $15\%$ for further testing. 
Preliminary experiments have been conducted to optimize the hyperparameters of the model, and are reported in the Appendix Table(\ref{tab:preliminaryCLIP}).
In Table(\ref{tab:CLIP}), we report the performances of the CLIP model averaged over various trials. The higher and more stable performances are found for the regression of the bulk modulus, but overall we observe promising generalization capabilities in terms of $R^2$, on average around $0.8$. We report examples of learning curves over different folds in the Appendix Fig.(\ref{fig:CLIP-curves1}). 
Additional experiments have been conducted by taking the Fourier Transform of ECD images to test the importance of periodicity. The Appendix Fig(\ref{fig:viridis-FT}) and Table(\ref{tab:preliminaryCLIP-FT}) respectively report examples of such data and the related performances of the model, which however did not underline specific improvements. 

\begin{figure}
    \centering
    \includegraphics[width=0.99\linewidth]{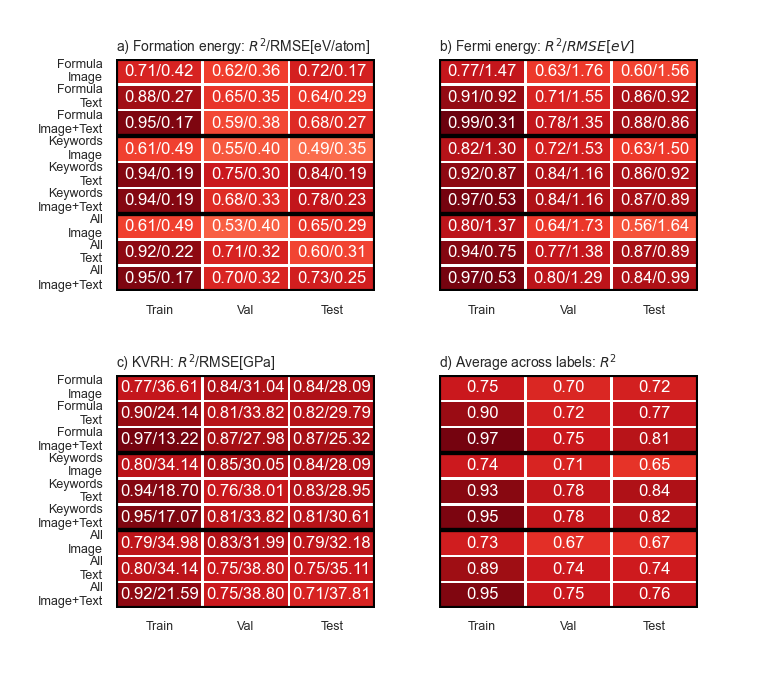}
    \caption{Regression performance of the different stages of a custom stacked multimodal model towards the three target properties: the formation energy, $E_F$, the Fermi energy, $E_{fermi}$, and the bulk modulus, $K_{vrh}$. The 'Image' only stage represents the fine-tuning (FT) of the InceptionV3 model. The 'Text' only stage represents the fine-tuning of the RoBERTa-Base model. The 'Image+Text' stage represents the fine-tuning of the embedding concatenation from the two modalities. Each of the stages involving text, also report the extent of textual information used, being it only the 'Formula', the 'Keywords' (composition name and bond lenghts information) or 'All' available text. }
    \label{fig:stackedperform}
\end{figure}

Fig.(\ref{fig:stackedperform}) reports the regression performance of the stacked custom multimodal model during the different stages of its implementation. Already at the stage of InceptionV3 (image-only) finetuning, the bulk modulus prediction reaches high $R^2$ values in generalization, comparable to the CLIP model, and the subsequent finetuning of the RoBERTa (text-only) model or the composition of the semantic and visual embeddings from the finetuned models do not seem to provide visibile improvements.  
The prediction of the Fermi energy based on ECD images only, instead, does not generalize well, and remarkably benefits from the use of textual crystallographic information at all its levels (Formula, Keywords and All). 
The formation energy is a challenging prediction target, though it shows slight improvement as text input is added.

%I should interpret these results as interepretability support for what happens in clip. The bulk modulus does not need text, the fermi energy needs a middle amount, the formation energy a large amount

\begin{table}[]
\caption{Statistical results for regression of the target properties with the CLIP multimodal model showing the mean R$^2$ across validation folds, as well as the test set R$^2$ for each run, along with the average values. CLIP used the full textual description available ('All'). }
\label{tab:CLIP}
\centering
\begin{tabular}{|c|c|c|c|c|} \hline 
&&\textbf{$E_F$}& \textbf{$E_{fermi}$}&\textbf{$K_{vrh}$}\\ \hline  
&& \text{R$^2$}/RMSE[eV/atom] & \text{R$^2$}/RMSE[eV] & \text{R$^2$}/RMSE[GPa] \\ \hline  
\multirow{2}{*}{\text{\#1}}& Val. Folds Average &0.91/0.21 & 0.90/0.89 & 0.91/21.9 \\ %\hline  
&Test &0.72/0.36 & 0.87/0.90 &0.88/24.4 \\ \hline  
\multirow{2}{*}{\text{\#2}}& Val. Folds Average &0.88/0.23 & 0.90/0.88& 0.88/24.1 \\ %\hline  
&Test &0.89/0.24 & 0.82/1.25&0.80/35.3 \\ \hline  
\multirow{2}{*}{\text{\#3}}& Val. Folds Average &0.77/0.35 & 0.90/0.86 &0.92/21.1 \\ %\hline  
&Test &0.67/0.35 & 0.51/2.07&0.80/36.5 \\ \hline  
\multirow{2}{*}{\textbf{$\langle \# \rangle$}}& \textbf{Val. Folds Average} &0.85/0.26& 0.9/0.86&0.90/22.4\\ %\hline  
& \textbf{Test set} &0.76/0.31& 0.73/1.41&0.83/32.1\\ \hline  
\end{tabular}
\end{table}

\section{Discussion, Limitations and Future Work}
In this work we propose computationally efficient approaches to property prediction in materials through the use of multiple modalities in transfer learning, with ECDs as highly informative input data. In particular, we show the dual approach of a ready-to-use pre-trained multimodal model like CLIP and the creation of a custom stack of single-modality pre-trained models. Even though the performances of the two approaches are similar, the latter stands as i) a more interpretable solution, because the single-modality stages convey the importance of each model and data type in the final combined performance; ii) more versatile solution, in it being composed of arbitrary single-modality models, which might be chosen appropriately for the planned application.
We underline the highly informative nature of the data: the image of a single slice out of a three-dimensional field could be enough to build expressive features for the prediction of computationally expensive material properties properties like the bulk modulus, which generally requires calculating the total energy of the system under different volumes and then fitting to an equation of state. 
Furthermore, we observe remarkable regression improvements through inclusion of semantic embeddings, as opposed to only using visual information. It is an encouraging sign of the possibilities that multimodal approaches offer in such context. 

The work presents some limitations and potential for future work. As we run independent trials for different variants of textual descriptions, we notice that multiple instances of image-only finetuning show palpable variance in performance, even though they are not influenced by varying text input (Fig. (\ref{fig:stackedperform}, Image-FT). This points at the need for multi-run experiments to quantify the variance in performance.
An exciting avenue already under consideration could be the implementation of multimodal feature estimation in an inverse design procedure, which will be able to employ ECD fields and basic crystallographic information in a search for desired set of properties. Such approach would represent a remarkable leap in the field, connecting low-cost approaches, generative models and ECD prediction.

\begin{ack}
The Author(s) D. Massa and S. Papanikolaou were supported by the European Union Horizon 2020 research and innovation program under Grant Agreement No. 857470 and from the European Regional Development Fund under the program of the Foundation for Polish Science International Research Agenda PLUS, grant No. MAB PLUS/2018/8, and the initiative of the Ministry of Science and Higher Education 'Support for the activities of Centers of Excellence established in Poland under the Horizon 2020 program' under agreement No. MEiN/2023/DIR/3795.
Piotr Sankowski was partially supported by ERC consolidator grant TUgbOAT No. 772346 and NCN No. 2020/37/B/ST6/04179.
\end{ack}

\bibliographystyle{unsrt}
\bibliography{biblio}
%%%%%%%%%%%%%%%%%%%%%%%%%%%%%%%%%%%%%%%%%%%%%%%%%%%%%%%%%%%%
\newpage
\appendix

\section{Appendix / supplemental material}

\begin{figure}[H]
    \centering
    \includegraphics[width=\linewidth]{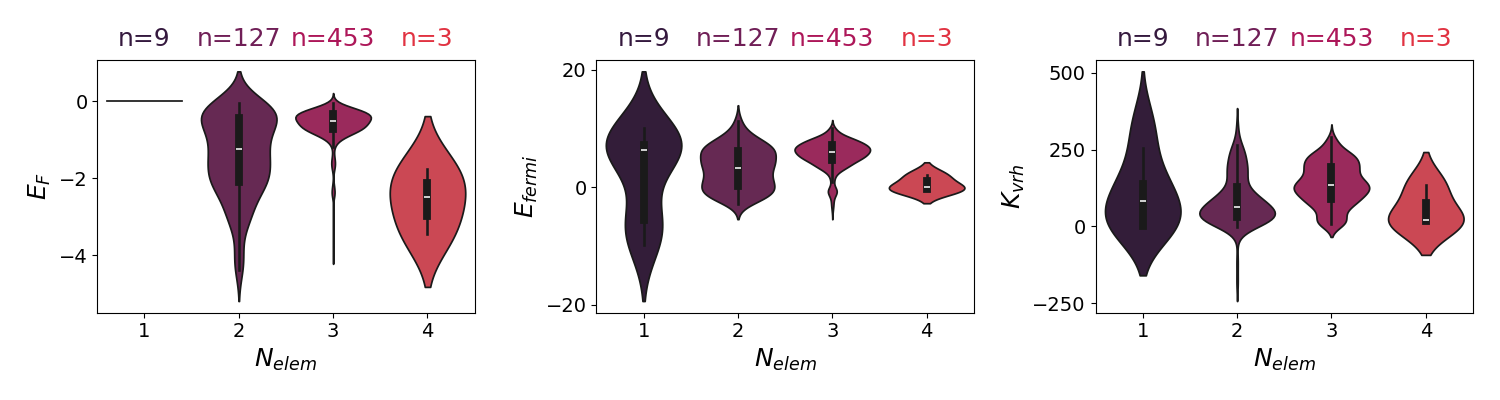}
    \caption{Visualization of the samples distributions according to the three target properties, the formation energy per atom ($E_F$ in eV/atom), the Fermi energy ($E_{fermi}$ in eV) and the bulk modulus ($K_{vrh}$ in GPa), and to the number of constitutive atomic species ($N_{elem})$. For each violin plot, a number $n$ on the upper edge of the frame indicates the corresponding number of samples. The plot clearly indicates the majority of samples being three-species compounds. A wider (narrower) horizontal section of a single distribution instead indicates more (fewer) data points are present at that value. The white mark indicates the median, while the black bar the interquartile (IQR) range. The distributions here shown correspond already to the cleaned version of the dataset, with 592 samples.}
    \label{fig:datadistrib}
\end{figure}

\begin{figure}[H]
    \begin{subfigure}[t]{\textwidth}
        \centering
        \includegraphics[width=\textwidth]{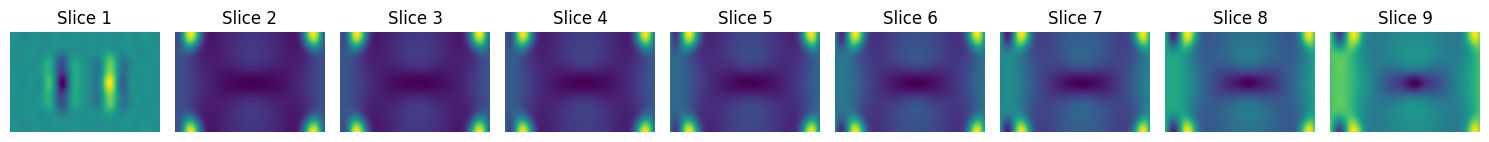}
%        \caption{Differential ECD slices}
        \label{fig:a}
    \end{subfigure}

    % Insert text here
    \begin{minipage}{\textwidth}
        \centering
        a) Differential ECD slices.
    \end{minipage}

    \begin{subfigure}[b]{\textwidth}
        \centering
        \includegraphics[width=\textwidth]{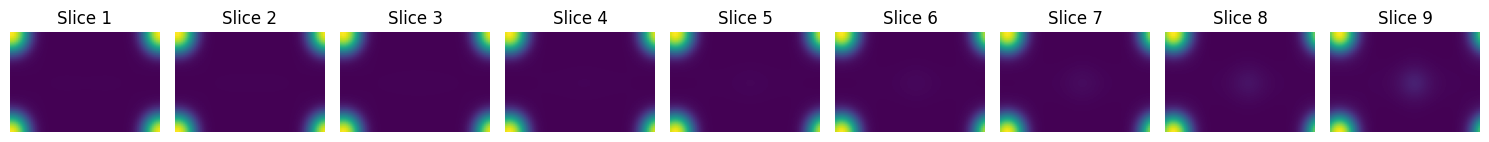}
%        \caption{Total ECD slices}
        \label{fig:b}
    \end{subfigure}
    
    % Insert text here
    \begin{minipage}{\textwidth}
        \centering
        b) Total ECD slices.
    \end{minipage}
    
    \caption{An example of 2D slices along the z-axis of the ECD volumetric data for an MP sample (mp-72, Ti).}
    \label{fig:zlices}
\end{figure}

\begin{figure}
    \centering
    \includegraphics[width=\linewidth]{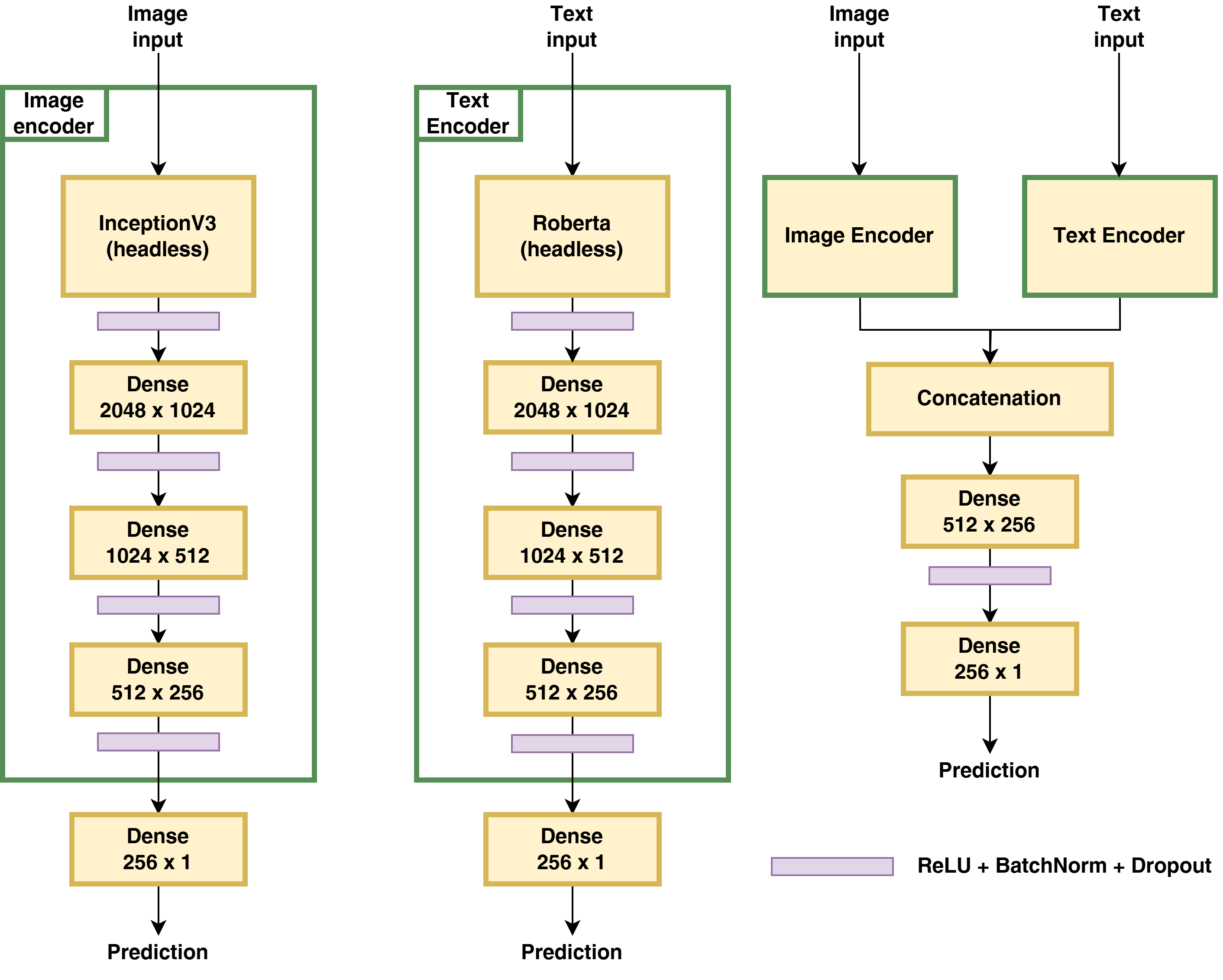}
    \caption{Schematic representation of the stacked custom multimodal model architecture. From the left: image-only model, text-only model, and a combined multimodal model that utilizes them both.}
    \label{fig:scheme}
\end{figure}

\begin{table}[H]
\caption{Different trials (T\#) for the hyperparameters used during the finetuning of the CLIP model and associated performance metrics. The images correspond to ECD images of the samples in viridis colorscale, while the target property for regression is always the formation energy.}
\label{tab:preliminaryCLIP}
\centering
\begin{tabular}{lccccccccc}
\hline
& BS & LR & Patience & Patch & $\text{Ratio}_\text{T}$ & $\langle \text{R}^2\rangle_\text{f}$ & $\langle \text{RMSE} \rangle_\text{f}$& $\text{R}^2_\text{T}$& $\text{RMSE}_\text{T}$ \\
\hline
T\#1 & 16 & 0.00015 & 17 & 32 & 0.15 & 0.6373 & 0.5873 & 0.8476 & 0.4127 \\
T\#2 & 32 & 0.00015 & 50 & 32 & 0.15 & 0.9064 & 0.2973 & 0.7792 & 0.4967 \\
T\#3 & 16 & 0.00010 & 50 & 32 & 0.15 & 0.8548 & 0.3670 & 0.8942 & 0.3438 \\
T\#4 & 32 & 0.00015 & 50 & 16 & 0.15 & 0.8298 & 0.4033 & 0.6508 & 0.6246 \\
T\#5 & 16 & 0.00010 & 50 & 16 & 0.15 & 0.6988 & 0.4990 & 0.7123 & 0.5670 \\
\hline
\end{tabular}
\end{table}

\begin{figure}[H]
    \centering
    \includegraphics[width=\linewidth]{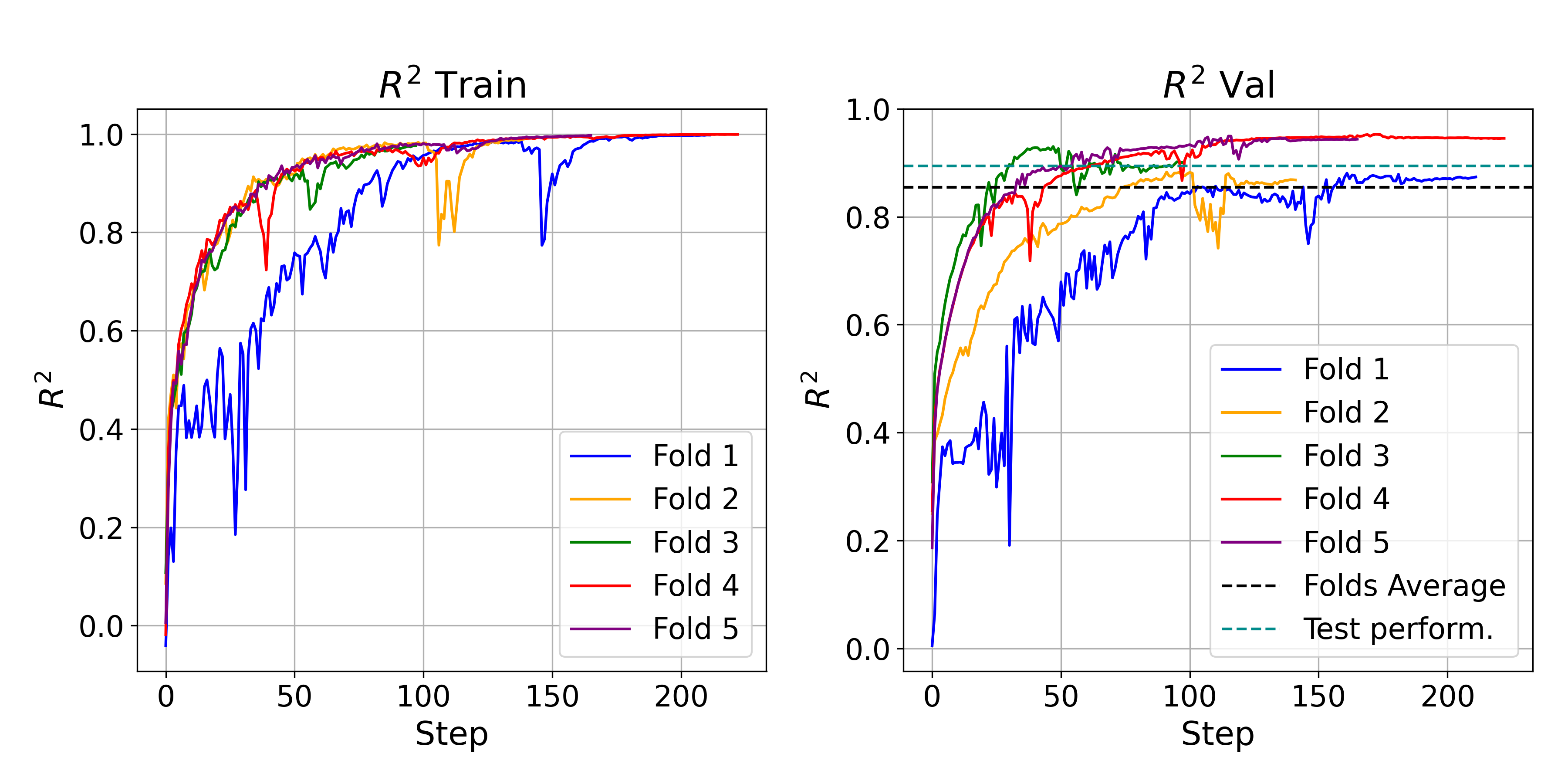}
    \caption{Performance of the selected hyperparameters set (T\#3) on the ECDs dataset composed of images in viridis scale, in the regression of the formation energy property.}
    \label{fig:CLIP-curves1}
\end{figure}
\begin{figure}[H]
    \centering
    \includegraphics[width=\linewidth]{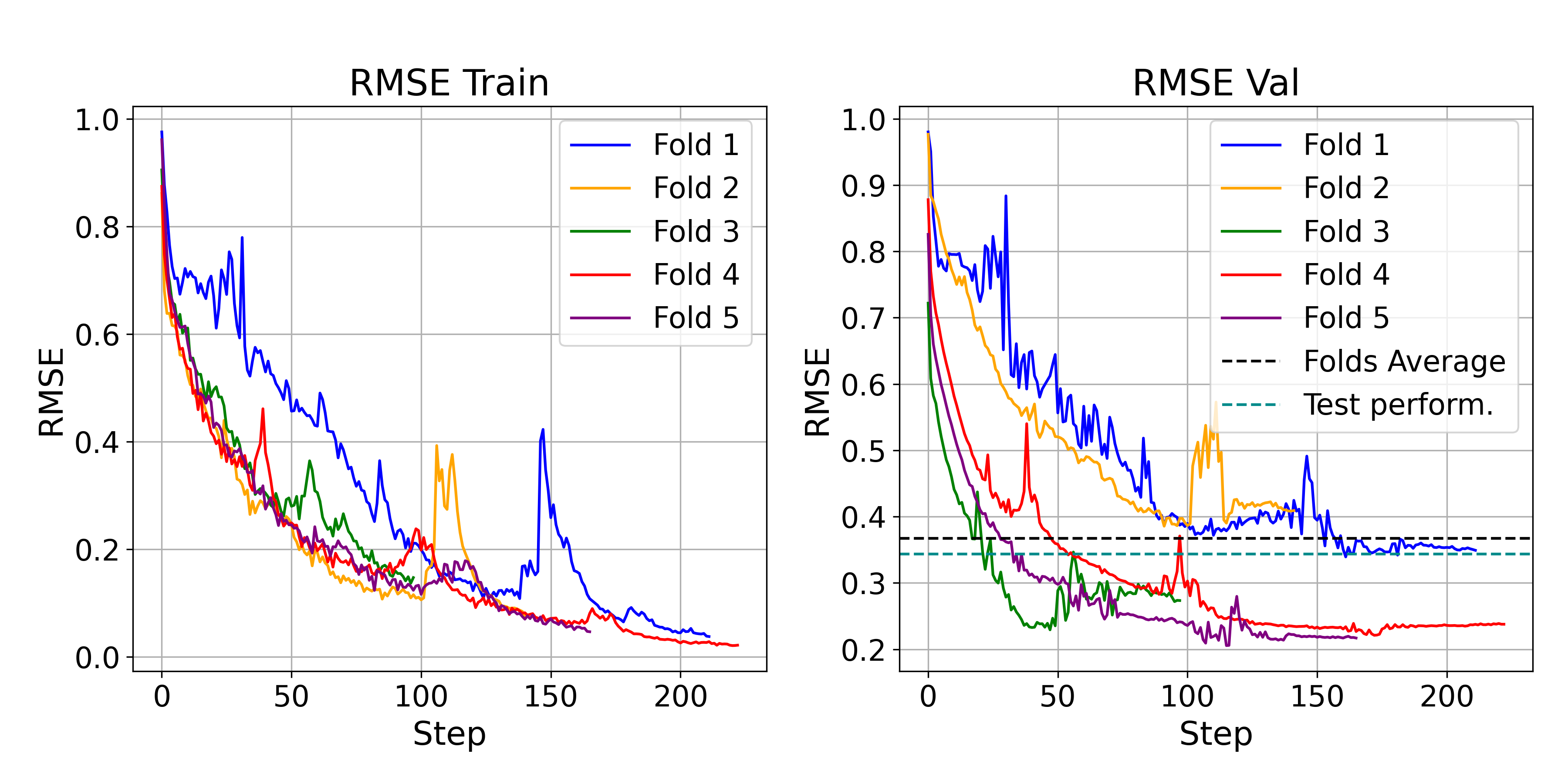}
    \caption{Performance of the selected hyperparameters set (T\#3) on the ECDs dataset composed of images in viridis scale, in the regression of the formation energy property. Being the target property standardized, the RMSE is adimensional. }
    \label{fig:CLIP-curves2}
\end{figure}

\begin{figure}[H]
    \centering
    \includegraphics[width=\linewidth]{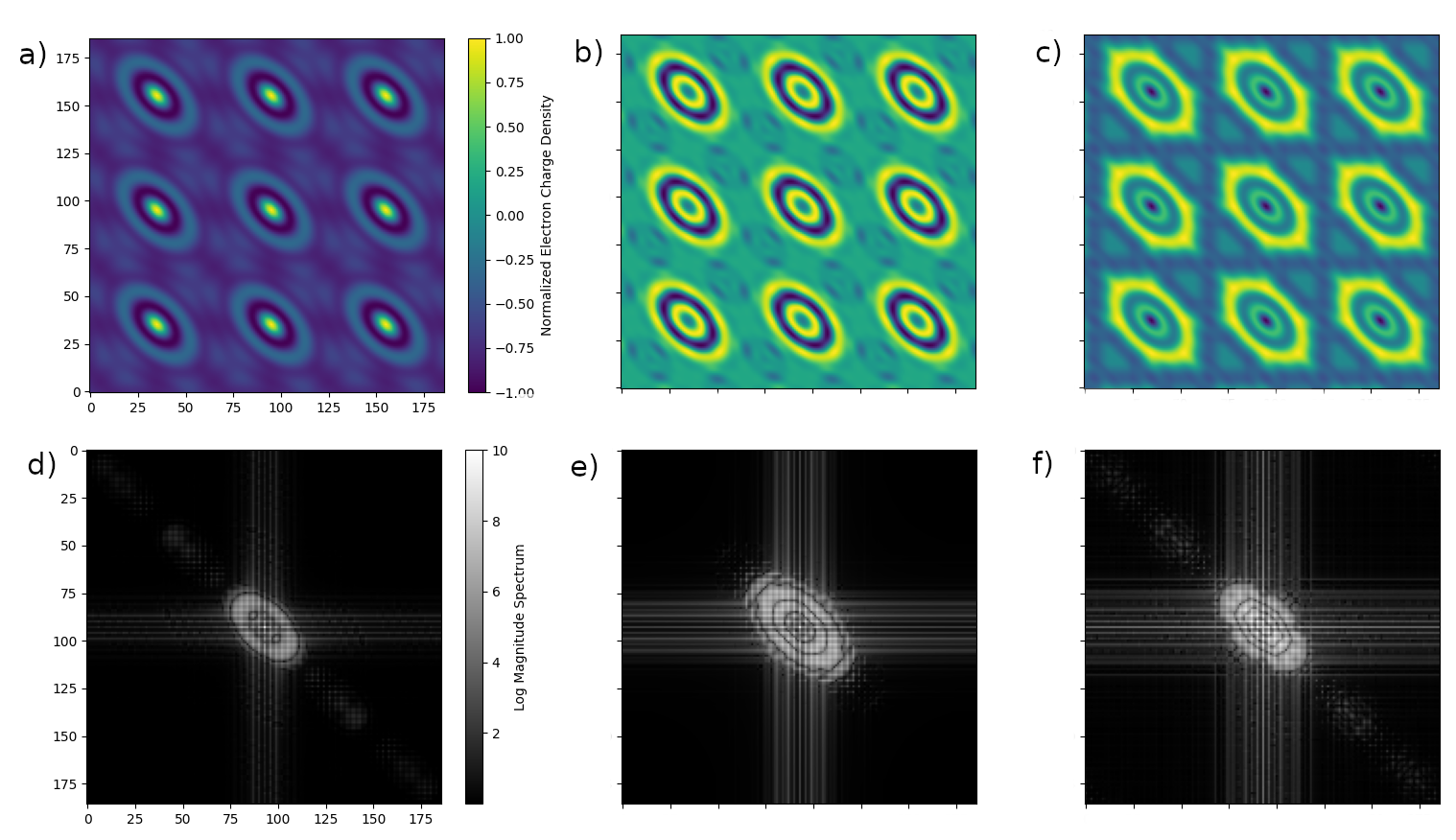}
    \caption{a-c) Visualization in viridis colorscale of three examples of periodic and normalized ECD patterns extracted from the center of the sliced planes. d-f) Fourier transforms of the upper ECD image data.}
    \label{fig:viridis-FT}
\end{figure}

\begin{figure}[H]
    \centering
    \includegraphics[width=\linewidth]{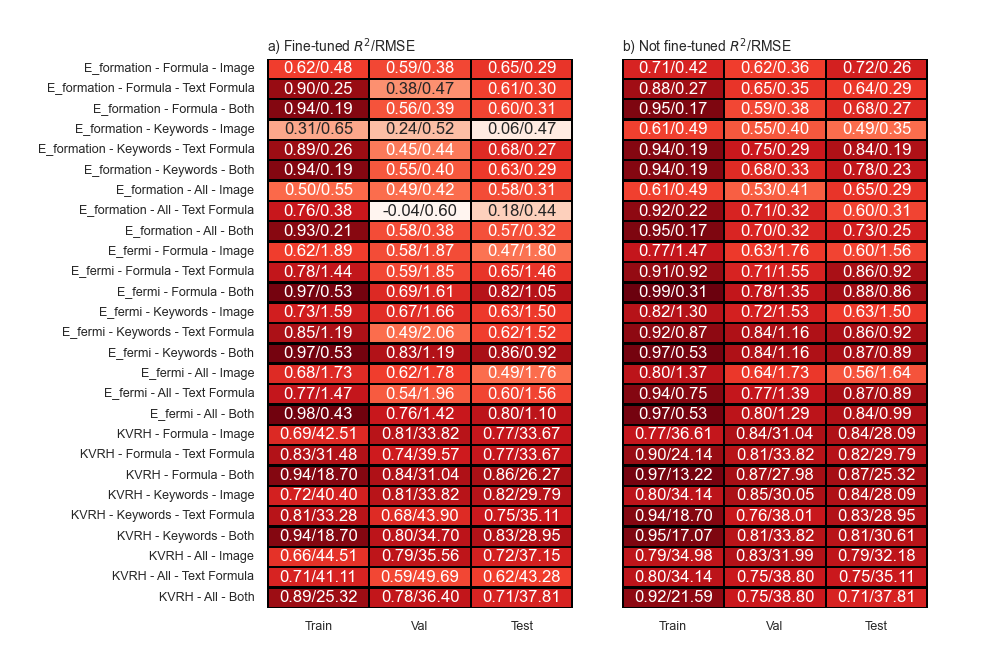}
    \caption{Performance of the models before and after fine-tuning.}
    \label{fig:viridis-FT}
\end{figure}

\begin{table}[H]
\caption{Different trials (T\#) for the hyperparameters used during the finetuning of the CLIP model and associated performance metrics. The images correspond to the Fourier transform of the ECD images in grey scale, while the target property for regression is always the formation energy.}
\label{tab:preliminaryCLIP-FT}
\centering
\begin{tabular}{lccccccccc}
\hline
& BS & LR & Patience & Patch & $\text{Ratio}_\text{T}$ & $\langle \text{R}^2\rangle_\text{f}$ & $\langle \text{RMSE} \rangle_\text{f}$& $\text{R}^2_\text{T}$& $\text{RMSE}_\text{T}$ \\
T\#1& 16 & 0.00015 & 17 & 32 & 0.15 & 0.43& 0.73& 0.65 & 0.62\\
T\#2& 32 & 0.00015 & 50 & 32 & 0.15 & 0.84& 0.38& 0.72& 0.56\\
T\#3& 16 & 0.00010 & 50 & 32 & 0.15 & 0.90& 0.29& 0.68& 0.60\\
T\#4& 32 & 0.00015 & 50 & 16 & 0.15 &  0.81& 0.41& 0.72& 0.56\\
T\#5& 16 & 0.00010 & 50 & 16 & 0.15 & 0.86& 0.35& 0.68& 0.60\\
\hline
\end{tabular}
\end{table}

\end{document}